\documentstyle[preprint,pre,aps]{revtex}
\tightenlines
\begin{document}



\title{Statistical mechanics of \\
image restoration and error-correcting codes}
\author{Hidetoshi Nishimori}
\address{Department of Physics, Tokyo Institute of Technology,
Oh-Okayama, Meguro-ku, Tokyo 152-8551, Japan}
\author{K. Y. Michael Wong}
\address{Department of Physics,
Hong Kong University of Science and Technology,
Clear Water Bay, Kowloon, Hong Kong}

\date{\today}

\maketitle

\begin{abstract}
We develop a statistical-mechanical formulation
for image restoration and error-correcting codes.
These problems are shown to be equivalent to the Ising spin
glass with ferromagnetic bias under random external fields.
We prove that the quality of restoration/decoding
is maximized at a specific set of parameter values determined
by the source and channel properties.
For image restoration in mean-field system
a line of optimal performance is shown to exist
in the parameter space.
These results are illustrated by solving exactly
the infinite-range model.
The solutions enable us to determine how precisely one should
estimate unknown parameters.
Monte Carlo simulations are carried out to see how far the
conclusions from the infinite-range model are applicable to the
more realistic two-dimensional case in image restoration.

\end{abstract}


\narrowtext
\section{INTRODUCTION}

Information is usually transmitted through noisy channels.
One therefore has to devise a method to retrieve the original
information from the output of a noisy channel.
Let us suppose that the original information is represented
as a sequence of bits.
The idea of error-correcting codes \cite{ecc} is to introduce
redundancy into the bit sequence to be fed into the noisy channel,
so that this additional information
is helpful to retrieve (decode) the original bit sequence
from the corrupted output of the channel.
An example is the parity-check code in which the
parities of appropriate blocks of bits are
sent through the channel in addition to the original
source sequence.
The receiver checks the consistency between the information
bits and parity bits and takes appropriate actions
if an inconsistency is found.

The problem of image restoration \cite{Pryce} is
similar to the error-correcting code, in the sense
that an image represented by
a set of pixels (corresponding to the bit sequence in
error-correcting codes) is corrupted by noise
and the receiver tries to retrieve the original image out of the
noisy, corrupted one.
A major difference is that in the image restoration problem,
one is usually given only the corrupted image,
not other additional redundant information.
One thus relies on some {\it a priori} knowledge about
images in general to remove noise.
A simple instance is the assumption of
smoothness in real-world images;
for example, one may wish to suppress an isolated white pixel
among black ones because such a configuration is likely to have
been caused by noise rather than to have existed
in the original image of the real world.

A general strategy common in error-correcting
codes and image restoration is to use the Bayes formula
on the {\it a posteriori} probability (posterior)
of an output sequence, given the input sequence.
One then often accepts the sequence (image) which maximizes
the posterior as the
decoded/restored result.
This method is called the maximum
{\it a posteriori} probability (MAP) estimate.

Sourlas \cite{Sourlas89} pointed out that the problem of
error-correcting codes can be written in terms of
the theory of spin glasses.
The idea is first to represent the bit sequence as
an Ising spin configuration and then to form a set of
exchange interactions as the products of appropriate
sets of spins, which can be considered as
a generalization of the Mattis model
of spin glasses \cite{Mattis}.
The set of interactions, instead of the spin configuration,
is fed into the channel, in
which noise causes the signs of interactions to flip
with some probability.
At the receiving end, one forms the Ising model Hamiltonian
from the corrupted exchange interactions, the ground state
of which is accepted as the decoded information.
This process is equivalent to the MAP estimate.
Sourlas used this formulation to show that there exists a
family of codes which are asymptotically error-free and
saturate the bound
on the code rate (the number of information bits
divided by the number of transmitted bits)
derived by Shannon \cite{Shannon}.

Ruj\'an \cite{Rujan} subsequently proposed to carry out
the decoding
procedure not at the ground state but at a finite
temperature corresponding to the Nishimori temperature
found in the theory of spin glasses \cite{Nishimori81}.
The finite-temperature decoding is effective
because the correct original bit sequence has a higher
energy than the ground state if the exchange interactions
are corrupted and thus are deviated from the Mattis type.
His proposal was supported by one of the authors
\cite{Nishimori93} who proved that
when the decoding temperature is varied,
the average error per bit in the decoded sequence
becomes smallest at the Nishimori temperature.
Sourlas \cite{Sourlas94} used the Bayes formula
to rederive the finite-temperature decoding
of Ruj\'an under more general conditions.
A recent development along this line is the
contribution by Kabashima and Saad \cite{Kabashima}
who used a diluted many-spin interacting model of spin
glasses to show that this system asymptotically saturates
the Shannon bound with the code rate kept
finite.

Analogy with statistical mechanics has also
been a useful guide
to develop a variety of techniques in the image restoration
problem \cite{Geman,Zerubia,TanakaPL}.
Most of the efforts, however, have been devoted to the development
of efficient methods to search for the ground state of
appropriate statistical mechanical systems
(the MAP restoration) using simulated annealing \cite{Geman},
mean-field annealing \cite{Zerubia} or the cluster variation
methods \cite{TanakaPL}.
An interesting exception is the work of Pryce and Bruce \cite{Pryce}
who pointed out that under a certain criterion,
the MAP estimate is outperformed by
the thresholded posterior mean (TPM) estimate \cite{Marroquin},
which is actually equivalent to
a finite-temperature decoding method.

It is therefore natural to formulate the problems of
error-correcting codes and image restoration within a unified
theoretical framework and apply various techniques developed
in the theory of disordered systems like spin glasses
and the random-field Ising model.
In this paper, we study the choice of parameters
for optimal performance with this approach.
For illustration, we introduce the infinite-range model,
by which it becomes possible to discuss analytically
the parameter dependence of the performance.

In Sec. II we give the basic formulation of the problem
for the binary symmetric and Gaussian channels.
The posterior for the output of a noisy channel is interpreted as
the Boltzmann factor of a statistical-mechanical system, namely
the Ising spin glass with ferromagnetic bias under random fields.
Important quantities in the theory of error-correcting codes and
image restoration are represented in terms of thermal averages
of the Ising model at finite temperatures.
We derive an upper bound
on the overlap between the decoded/restored result
with the original sequence/image
using the statistical-mechanical formulation.
The problem of image restoration is treated in Sec. III in detail 
where we derive a line of optimal performance in mean-field systems, 
find the exact solution for the infinite-range model, 
and present simulation results for the two-dimensional case.
The infinite-range model is shown to work as a good guide
to the description of qualitative behavior of the two-dimensional
problem.
Explicit examples of images are displayed to clarify what happens
under various conditions.
The problem of error-correcting codes is analyzed
in Sec. IV.
We solve the infinite-range model explicitly using
the replica method.
Compact expressions of the overlap and other order parameters
enable us to discuss various aspects of error-correcting
codes quantitatively.
%
%
The final section is devoted to discussions.
%

\section{GENERAL FORMULATION}

Consider an information source which generates a bit sequence
represented by a set of Ising spins $\{ \xi_i\}$,
where $\xi_i=\pm 1$ and $i=1,\cdots ,N$,
with the source probability $P_s(\{\xi_i\} )$ (the prior).
The sequence $\{\xi_i\}$ is coded as the products of $r$ spins
$J^0_{i_1\cdots i_r}=\xi_{i_1}\cdots\xi_{i_r}$
for appropriately chosen sets of indices
$\{ i_1\cdots i_r \}$.
The Sourlas code \cite{Sourlas89} is equivalent to the
infinite-range model in which all possible combinations
of $r$ sites are chosen from $N$ sites.
In general we consider several different $r$'s in a
single code, such as in the case of the Viterbi code
\cite{Rujan} which has $r=2$ and 3.
The problem of image restoration can be regarded as a special
case of $r=1$, in which case $J_i$ corresponds to the state of the
$i$th pixel in the corrupted (noisy) image.
A ``bit" in error-correcting codes should thus be identified with
a ``pixel" in image restoration, a ``bit sequence" with an ``image",
and ``decoding" with ``restoration"
whenever necessary in the following arguments.

When the signal is transmitted through a noisy channel,
the output consists of the sets
$\{ J_{i_1\cdots i_r}\}$ for various values of $r$,
which are the corrupted versions of $J^0_{i_1 \cdots i_r}$.
Two kinds of noisy channels are considered in this paper:
the binary symmetric channel (BSC) and the Gaussian channel (GC).
%

\subsection{Binary symmetric channel}

In the binary symmetric channel,
the output $J_{i_1 \cdots i_r}$ is equal to
$\mp J^0_{i_1 \cdots i_r}$ with probabilities
$p_r$ and $1-p_r$ respectively,
where $p_r$ is the error rate of the BSC for transmission of
$J^0_{i_1 \cdots i_r}$.
The error probabilities of flipping the signal $+1$ to $-1$
and $-1$ to $+1$ are the same.
This output probability can be written in a compact form as
\begin{equation}
 P_{\rm out}(\{ J\} |\{\xi\} )=
 \prod_r (2\cosh \beta_r)^{-N_r}
  \exp \left(\sum_r \beta_r \sum J_{i_1 \cdots i_r}\xi_{i_1}
   \cdots \xi_{i_r}\right) ,
   \label{BSC}
\end{equation}
where $\beta_r=0$ if the code in consideration does not include
the set $\{ J_{i_1 \cdots i_r}\}$ and 
\begin{equation}
        \beta_r =\frac{1}{2}\ln\frac{1-p_r}{p_r}
\label{betar}
\end{equation}
otherwise.
The second summation in the exponent of Eq. (\ref{BSC}) extends
over an appropriate set of the indices $(i_1, \cdots ,i_r)$,
the choice of which determines the type of the code, and
$N_r$ is the number of terms appearing in this summation.
Note that $N_r=0$ if $\beta_r=0$.
Each index $i$ may appear in a number of $r$-spin terms
in the exponential expression;
the number of times of appearance is called the valency $z_r$.

The procedure of decoding/restoration proceeds as follows.
According to the Bayes formula and Eq. (\ref{BSC}), the 
posterior probability that the source sequence is
$\{\sigma\}$, given the output $\{ J\}$, is
proportional to the Boltzmann factor of an Ising model
multiplied by the prior:
\begin{equation}
 P(\{\sigma\} |\{ J\} )\propto
  \exp \left(\sum_r \beta_r \sum J_{i_1 \cdots i_r}
           \sigma_{i_1}\cdots \sigma_{i_r} 
   \right) P_s(\{ \sigma\} ).
   \label{Bayes0}
\end{equation}
Note that we use the symbol $\{\sigma \}$ for the decoded/restored
result which is in general different from the original set $\{\xi \}$.

For simplicity, we restrict ourselves
to the case of a single non-vanishing $\beta_r(\equiv \beta_J)$
with $r\ge 2 $ 
and $\beta_1(\equiv \beta_\tau )$.
Following the convention of separating the interaction terms
and local field terms in statistical mechanics,
we write $J_i$ as $\tau_i$ for $r=1$ terms, 
and Eq. (\ref{Bayes0}) becomes
\begin{equation}
 P(\{\sigma\} |\{ J\} ,\{\tau \})\propto
  \exp \left(\beta_J \sum J_{i_1 \cdots i_r}
           \sigma_{i_1}\cdots \sigma_{i_r} 
           +\beta_\tau \sum \tau_i \sigma_i
   \right) P_s(\{ \sigma\} ),
   \label{Bayes}
\end{equation}
Similarly, we will use $p_J$, $p_\tau$
to represent $p_r$, $p_1$ respectively.

It often happens that the receiver at the end of the
noisy channel does not have precise information on $\beta_J$,
$\beta_\tau$ or $P_s$.
One has to estimate these so-called hyperparameters,
which is one of the major topics in this field
(see for example \cite{Zhou}).
If the receiver estimates $\beta$ for $\beta_J$ and $h$
for $\beta_\tau$,
and uses a model (guess) $P_m(\{\sigma\})$
of the source prior $P_s(\{\xi\})$,
then the mean of the posterior distribution of $\sigma_i$
is equal to the thermal average
\begin{equation}
 \langle \sigma_i \rangle =
  \frac{ \sum_\sigma \sigma_i e^{-\beta H} P_m(\{\sigma\} )}
       { \sum_\sigma e^{-\beta H} P_m(\{\sigma\} )},
  \label{sigmai}
\end{equation}
where the Hamiltonian is given by
\begin{equation}
  \beta H=-\beta \sum J_{i_1 \cdots i_r}\sigma_{i_1}
         \cdots \sigma_{i_r}
   -h\sum \tau_i \sigma_i.
   \label{Hamiltonian}
\end{equation}
One then regards ${\rm sgn}\langle\sigma_i\rangle$ as the
$i$th bit of the decoded/restored information
with finite $\beta$ and $h$ in the finite-temperature process,
or with their ratio kept finite
when their magnitudes approach infinity in the MAP method.

In the context of image restoration, one often considers patterns
with non-trivial structures.
Therefore they are assumed to be generated by
a non-uniform source prior $P_s(\{\xi \})$.
When we do not have any information on the source prior,
we have to represent our {\it a priori} knowledge on
general images in the model prior $P_m(\{\xi \})$.
A natural choice often used as a generic form of the
prior is the Boltzmann factor of the ferromagnetic Ising model
\begin{equation}
 P_m(\{\sigma\} ) =
 \frac{1}{Z(\beta_m)}
        \exp \left({\beta_m\over z} \sum_{\langle ij\rangle}
       \sigma_i \sigma_j \right) ,
        \label{pm}
\end{equation}
where $Z(\beta_m)$ is the partition function
at the inverse temperature $\beta_m$,
$\langle ij\rangle$ represents interacting sites,
and $z$ is the valency of each site.
The summation usually extends over neighboring sites
on a two-dimensional lattice.
This prior is natural because it suppresses different states
of neighboring sites, enhancing a smooth structure.
In this paper we consider priors with general connections.

To proceed further, we have to assume some explicit form of
the source prior $P_s(\{\xi \})$.
To develop a general theory,
we adopt the Boltzmann factor of the Ising model
\begin{equation}
 P_s(\{\xi\} ) = 
   \frac{1}{Z(\beta_s)}
        \exp \left({\beta_s\over z}\sum_{\langle ij\rangle}
         \xi_i \xi_j \right),
        \label{ps}
\end{equation}
which has the same form as Eq. (\ref{pm})
but with a different inverse temperature.
Thus the original images correspond to snapshots of equilibrium
Monte Carlo simulations of the ferromagnetic Ising model.

Comparison of Eqs. (\ref{Bayes}), (\ref{sigmai}) and
(\ref{Hamiltonian}) implies that the Bayes result (\ref{Bayes})
specifies the inverse temperature $\beta$ to $\beta_J$,
the field strength $h$ to $\beta_\tau$ and
the model prior $P_m$ to $P_s$.
Nevertheless it is useful to keep $\beta$, $h$ and $P_m$
as adjustable parameters and investigate how precisely
we should estimate the hyperparameters and tune the adjustable
parameters for optimal decoding/restoration.
Our statistical-mechanical formulation is particularly
useful to investigate this problem both qualitatively
and quantitatively.

The decoding/restoration procedure
reduces to the method of Ruj\'an in the
situation of error-correcting codes ({\it i.e.} $h=0$ and
$P_s=P_m=$const) when $\beta=\beta_J$, in which case the
noise temperature $1/\beta_J$ is equal to the Nishimori
temperature, which is known to play an interesting role
in the theory of spin glasses \cite{Nishimori81}.
Pryce and Bruce \cite{Pryce} called
the same method the TPM in the image restoration situation,
in which one sets $h=\beta_\tau$ in the absence of
the interaction term in Eq. (\ref{Hamiltonian}).

The finite-temperature method with appropriate parameters
($\beta =\beta_J, h=\beta_\tau, P_m=P_s$)
is known to give the sequence of most probable
bits both in error-correcting codes and image restoration
\cite{Pryce,Nishimori93,Sourlas94,Marroquin}.
The MAP estimate, on the other hand, chooses the sequence which
gives the largest value of the posterior (\ref{Bayes}),
corresponding to the ground state of the Ising model.
The result of finite-temperature decoding/restoration gives
a lower value of the posterior (\ref{Bayes}) than the
MAP result.
However, there appear very many states with almost the same
value of the posterior if we generate finite-temperature
states by Eq. (\ref{Bayes}), and thus after weighted by
the number of such similar states,
finite-temperature states outweigh the MAP counterpart.
In other words, if we take into account the entropy effects,
the finite-temperature method becomes the natural choice.
This corresponds to the free-energy minimization rather
than the energy minimization as in usual statistical-mechanical
systems at finite temperatures.
%

\subsection{Gaussian channel}

In the Gaussian channel, the output $J_{i_1 \cdots i_r}$
is a Gaussian random variable with appropriate
mean $a_r J^0_{i_1 \cdots i_r}$ and variance $J_r^2$.
Hence for a given sequence $\{\xi_i\}$,
the Gaussian channel is given as
\begin{eqnarray}
  &&P_{\rm out}(\{ J\},\{\tau \} |\{\xi\} ) 
        \propto
     \exp \left( -\frac{1}{2J^2}
      \sum (J_{i_1 \cdots i_r}-J_0 \xi_{i_1}\cdots \xi_{i_r})^2
       -\frac{1}{2\tau^2}
       \sum (\tau_i -a \xi_i )^2\right)
       \label{gauss1}
       \\
    &&=\exp \left( -\frac{1}{2J^2}
      \sum (J_{i_1 \cdots i_r}^2+J_0 ^2)
        -\frac{1}{2\tau^2}
       \sum (\tau_i^2 +a^2)
       +\frac{J_0}{J^2}\sum J_{i_1 \cdots i_r}
      \xi_{i_1}\cdots \xi_{i_r}
      +\frac{a}{\tau^2}\sum \tau_i \xi_i \right),
    \label{gauss2}
\end{eqnarray}
where, again following the statistical mechanics convention,
we have used $J_0$, $J$, $a$, $\tau$ to represent
$a_r$, $J_r$, $a_1$, $J_1$ respectively.

Similarity of Eqs. (\ref{Bayes}) and (\ref{gauss2}) implies
that the output probability distribution for BSC and GC 
can be written in the same form,
\begin{equation}
 P_{\rm out}(\{ J\} |\{\xi\} )=
  \prod F_r(J_{i_1 \cdots i_r})\prod F_1(\tau_i)
  \exp \left(\beta_J\sum J_{i_1 \cdots i_r}
  \xi_{i_1}\cdots \xi_{i_r}+
  \beta_\tau\sum \tau_i\xi_i \right),
   \label{general}
\end{equation}
where $\beta_J=J_0/J^2$ and $\beta_\tau=a/\tau^2$ for GC.
$F_r$ and $F_1$ are functions independent of $\{\xi_i\}$.
For BSC,
\begin{eqnarray}
        F_r(J_{i_1 \cdots i_r})
        &=&\frac{1}{2\cosh\beta_J}\left\{
        \delta(J_{i_1 \cdots i_r}-1)+
        \delta(J_{i_1 \cdots i_r}+1)\right\},\\
        F_1(\tau_i)
        &=&\frac{1}{2\cosh\beta_\tau}\left\{
        \delta(\tau_i-1)+\delta(\tau_i+1)\right\}.
\end{eqnarray}
For GC,
\begin{eqnarray}
        F_r(J_{i_1 \cdots i_r})
        &=&\frac{1}{\sqrt{2\pi J^2}}\exp\left(
        -\frac{1}{2J^2}(J_{i_1 \cdots i_r}^2+J_0 ^2)\right),\\
        F_1(\tau_i)
        &=&\frac{1}{\sqrt{2\pi \tau^2}}\exp\left(
        -\frac{1}{2\tau^2}(\tau_i^2 +a^2)\right).
\end{eqnarray}
%

\subsection{Overlap}

The most important quantity in the present problem is
the overlap of the decoded/restored bit
${\rm sgn}\langle \sigma_i \rangle$
and the original bit $\xi_i$
averaged over the output probability.
We may express this overlap as
\begin{equation}
   \prod \int dJ \prod \int d\tau
        P_{\rm out}(\{ J\},\{\tau \}|\{ \xi \} )
   ~ \xi_i \, {\rm sgn}\langle \sigma_i \rangle .
   \label{average1}
\end{equation}
This expression (\ref{average1}) should be further averaged over
the possible sequences of source bits represented by the
prior $P_s(\{\xi\} )$.
The final expression of the overlap $M$ is then
\begin{equation}
 M(\beta, h, P_m)=
   \sum_\xi \prod \int dJ \prod \int d\tau
   P_s(\{\xi\} ) P_{\rm out}(\{ J\},\{\tau \} |\{\xi \} )
   \, \xi_i \, {\rm sgn}\langle \sigma_i \rangle .
   \label{M}
\end{equation}
The dependence of $M$ on $\beta , h$ and $P_m$ exists in the
thermal average $\langle \sigma_i \rangle$.
The average of any other quantity $f(\sigma )$ is calculated
similarly:
\begin{equation}
 [\langle f\rangle ]=
   \sum_\xi \prod \int dJ \prod \int d\tau
     P_s(\{\xi\} ) P_{\rm out}(\{ J\},\{\tau \}|\{\xi \} )
   \frac{\sum_\sigma f(\sigma ) e^{-\beta H} P_m(\{\sigma\} )}
        {\sum_\sigma e^{-\beta H} P_m(\{\sigma\} )}.
     \label{expectation}
\end{equation}
The outer brackets $[\cdots ]$ in Eq. (\ref{expectation})
denote the averages over
$\{ \xi\}, \{ J\}$ and $\{\tau \}$ with the weight
$P_s P_{\rm out}$.

The following inequality on the overlap is very useful
in discussions on the decoding/restoration performance,
and the proof is outlined in Appendix A:
\begin{equation}
  M(\beta, h, P_m)\le M(\beta_J, \beta_\tau, P_s).
  \label{bound1}
\end{equation}
This inequality means that the overlap becomes largest
when $\beta =\beta_J$, $h=\beta_\tau$ and $P_m=P_s$.
In this sense
the performance of the MAP corresponding to the limit
$\beta , h\to\infty$ cannot exceed that of
the finite-temperature decoding/restoration
at $\beta_J$ and $\beta_\tau$.

This inequality was known in error-correcting codes
\cite{Nishimori93,Sourlas94}, and
equivalent statements were given also in the image
restoration problem \cite{Marroquin} although
not in the explicit form given here.
One of the contributions of the present paper is that
we have generalized the inequalities to the case where
two different types of terms are present,
one with $r\ge 2$ and the other with $r=1$.
Similar inequalities can be easily derived for the more
general case of several different $r$'s in Eq. (\ref{BSC}).
%

\section{IMAGE RESTORATION}
\subsection{Line of optimal performance in mean-field models}

In conventional image restoration problems,
the output of the transmission channel only consists
of the set of pixels $\{\tau_i \}$
(corrupted image in the usual sense),
but not the set of exchange interactions $\{J_{ij}\}$
(the corrupted version of $\{J_{ij}^0=\xi_i \xi_j\}$).
In this case the following inequality
applies to images with extensive valency.
Mean-field results are exact when each pixel interacts extensively
with other pixels in the source and model prior distributions
(\ref{ps}) and (\ref{pm}).
These include the infinite-range model,
in which all pixels interact with each other,
its randomly diluted version with infinite valency
in the thermodynamic limit ($N\to\infty$),
or finite-dimensional models with long-range interactions.
Mean-field approximations also work well
for finite but large valencies
when the temperature is not too low.
The following inequality derived in Appendix B is useful in
finding the optimal restoration performance,
\begin{equation}
  M(h, \beta_m)\le
  M\left(\beta_\tau s, \frac{\beta_s m_0}{m}s\right)
  \label{bound2}
\end{equation}
for arbitrary values of $s>0$.
Here $m_0$ and $m$ are the (self-averaging) thermal averages
of the source pixels $\xi_i$ and
model pixels $\sigma_i$ respectively,
for those sites $i$ which interact with a given site,
say site 1, which however is removed from the thermal process.
Note that the derivation only makes use of
the self-averaging nature of the mean-field quantities,
and does not rely on any connection topology of the sites
(except for the mean-field requirement).
No particular techniques
such as the replica method are employed.
Hence the inequality applies to mean-field systems in general.

Since $m_0$ and $m$ are functions of $\beta_s$, $\beta_m$ and $h$,
Eq. (\ref{bound2}) defines the line of optimal performance
in the space of $h$ and $\beta_m$:
\begin{equation}
  \frac{h}{\beta_\tau}=
  \frac{\beta_m m(\beta_s, \beta_m,h)}
  {\beta_s m_0(\beta_s)}=
  s.
  \label{line}
\end{equation}
In particular, when $s=1$,
$h=\beta_\tau$, $\beta_m=\beta_s$ and $m=m_0$,
Eq. (\ref{line}) reduces to the point of optimal performance
predicted by Eq. (\ref{bound1}).
On the other hand, if the field $h$ is different from $\beta_\tau$,
then the source and model temperatures have to be rescaled
by the magnetization of the respective systems.
When $s\to\infty$, we obtain the zero-temperature restoration,
i.e. the MAP estimate.
Hence the MAP estimate is also optimal,
provided that the correct ratio of $h/\beta_m$ is used
(although this choice can only be determined iteratively).
%

\subsection{The infinite-range model}

Let us now suppose that we are given the corrupted version
of the set of pixels $\{\tau_i \}$
and, in addition, the exchange interactions $\{J_{ij}\}$,
the latter being the corrupted version of the Mattis-type
interactions $\{\xi_i \xi_j\}$.
The additional contribution from the exchange term enables
us to formulate the image restoration and error-correcting
code problems on the same footing and at the same time
investigate the extent to which the additional information
$\{J_{ij}\}$ enhances the performance.
The restoration process is carried out at the inverse
temperature $\beta$
and the field $h$ using Eqs. (\ref{sigmai}) and
(\ref{Hamiltonian}) with $r=2$.
Though the inequalities (\ref{bound1}) and (\ref{bound2})
give a bound on the overlap $M$, they do not reveal
the explicit dependence of $M$
on the parameters $\beta, h$, $\beta_m$ and $\beta_\tau$,
which is necessary for studying
the tolerance of the restoration results
against uncertainties in the estimation of the hyperparameters.
The infinite-range model serves as a useful test ground for
this purpose.
Of course the infinite-range model is not useful for restoration
of a real two-dimensional image since all pixels are neighbors
of each other and hence the spatial structure is ignored.
However, we may reasonably expect from experience in statistical
mechanics of many-body systems that the behavior
of {\it macroscopic} quantities (such as the overlap $M$)
of realistic problems are at least qualitatively well predicted
by the infinite-range model.

We therefore suppose that the summation in the prior
(\ref{ps}) extends over all possible pairs of sites
\begin{equation}
 P_{s}(\{\xi\} ) = \frac{1}{Z(\beta_{s})}
 \exp \left( \frac{\beta_{s}}{2N} \sum_{i\ne j} \xi_i \xi_j \right) ,
\end{equation}
and similarly for the model prior $P_m$.
We also assume that the two-body exchange interactions for all
pairs of sites are included in the given information,
or equivalently Eq. (\ref{Hamiltonian}) with $r=2$ and the
summation extending over all pairs of sites,
and that the channel is Gaussian.
We thus have to evaluate the following averaged
replicated partition function
\begin{eqnarray}
  [Z^n]&=&
  \sum_\xi
    \int \prod_{i<j} \frac{dJ_{ij}}{\sqrt{2\pi J^2/N}}
     \exp \left( -\frac{N}{2J^2}
      \sum_{i<j} (J_{ij}-\frac{J_0}{N} \xi_{i}\xi_{j})^2
       \right)  \nonumber \\
   && ~~~\times
     \int \prod_i \frac{d\tau_i}{\sqrt{2\pi \tau^2}}
      \exp \left( -\frac{1}{2\tau^2}
       \sum_i (\tau_i -a \xi_i )^2 \right)
   \nonumber \\
   && ~~~\times \frac{1}{Z(\beta_s)}
 \exp \left( \frac{\beta_s}{2N} \sum_{i\ne j} \xi_i \xi_j \right)
    \nonumber \\
   && ~~~\times \sum_\sigma
   \exp \left(\beta\sum_{i<j}J_{ij}\sum_{\alpha =1}^n
             \sigma_i^\alpha \sigma_j^\alpha
           +\frac{\beta_m}{N}\sum_{i<j}\sum_{\alpha =1}^n
             \sigma_i^\alpha \sigma_j^\alpha
           +h\sum_i \tau_i\sum_{\alpha =1}^n \sigma_i^\alpha
           \right) .
 \label{IRZ}
\end{eqnarray}
The standard replica calculation with the replica symmetric ansatz
\cite{Fischer} leads to the expressions of the order parameters:
\begin{eqnarray}
 \left[\xi_i\right] &=& m_0=\tanh \beta_s m_0 \nonumber\\
 \left[\langle \sigma_i \rangle\right]
  &=&m=
     \frac{1}{2\cosh \beta_s m_0} \sum_{\xi =\pm 1}
     e^{\beta_s m_0 \xi} \int Dx \tanh U 
     \nonumber\\
 \left[\xi_i \langle \sigma_i \rangle \right] &=&t=
     \frac{1}{2\cosh \beta_s m_0} \sum_{\xi=\pm 1}
     \xi e^{\beta_s m_0 \xi} \int Dx \tanh U
     \nonumber\\
 \left[\langle \sigma_i \rangle^2 \right] &=&q=
     \frac{1}{2\cosh \beta_s m_0} \sum_{\xi=\pm 1}
     e^{\beta_s m_0 \xi} \int Dx \tanh^2 U ,
     \label{IReqs}
\end{eqnarray}
where
\begin{equation}
  U = \left(\beta^2 J^2 q+\tau^2 h^2\right)^{1/2}x +\beta_m m
      +(a h +\beta J_0 t )\xi .
\end{equation}
The overlap is a function of these order parameters
\begin{equation}
 \left[\xi_i {\rm sgn}\langle \sigma_i \rangle \right]=M
 = \frac{1}{2\cosh \beta_s m_0} \sum_{\xi=\pm 1}
    \xi e^{\beta_s m_0 \xi} \int Dx \, {\rm sgn} \, U \, .
    \label{Mi}
\end{equation}

An example of the dependence of $M$
on the model prior temperature $T_m\equiv\beta_m^{-1}$
is shown in Fig. \ref{fig_IR1}
for conventional image restoration
(without the exchange term, $\beta=0$).
The parameters are $T_s=0.9$ and $a=\tau=1$.
The usual practice in image restoration is to use a Hamiltonian
with a fixed ratio of $h/\beta_m$, and then to use $\beta_m$
as an adjustable parameter for simulated annealing.
Hence we consider the behavior as a function of $\beta_m$
when $h/\beta_m$ is kept constant to the optimal value
$\beta_\tau/\beta_s$ (the curve marked `Opt $h$'),
to 0.9 times the optimal value (`Opt*0.9'),
or to 1.1 times the optimal value (`Opt*1.1').
In the curve marked `$h=1$', $h$ itself is
kept constant to 1.
The ground-state limit $T_m\to 0$ gives the MAP restoration.
The maximum is at $T_m=0.9(=T_s)$ for the optimal choice of
$h=1(=\beta_\tau \beta_m/\beta_s)$,
as predicted by Eq. (\ref{bound1}).
This figure indicates that one does not have to approach
the zero temperature limit as in MAP
in the search of the best restored
image by such a process as simulated annealing.

For the curves Opt*0.9 and Opt*1.1,
the location of the largest $M$ is not at $T_m=0.9$.
However, the maximum value coincides with the best value
as predicted by Eq. (\ref{bound2}).
This fact can easily be verified by differentiating
Eq. (\ref{Mi}) with respect to $\beta_m$.
The optimal parameter is $\beta_m=\beta_s m_0 h/\beta_\tau m$,
agreeing with Eq. (\ref{line}).

The line of optimal performance in the space of $T_m$ and $h$ is
obtained by combining Eqs. (\ref{line}) and (\ref{IReqs}), yielding
\begin{equation}
        T_m=
     \frac{\beta_\tau}{\beta_s m_0 h}
        \frac{1}{2\cosh \beta_s m_0} \sum_{\xi =\pm 1}
     e^{\beta_s m_0 \xi} \int Dx \tanh \left(
        h\left(\frac{\beta_s m_0}{\beta_\tau}+a\xi+\tau x
        \right)\right).
\end{equation}
When $h\to 0$, $T_m$ approaches the limit
$(T_{m})_{\max} =1+a\beta_\tau/\beta_s$.
This is the temperature above which the maximum overlap
cannot be achieved.

In the low temperature limit, $h\to\infty$ and the ratio $h/\beta_m$
approaches a constant
\begin{equation}
        \lim_{h\to\infty}\frac{h}{\beta_m}=
     \frac{\beta_\tau}{\beta_s m_0} \sum_{\xi =\pm 1}
     \frac{1}{2}\left(1+m_0\xi \right)
        {\rm erf}\left(
        \frac{1}{\sqrt{2}\tau}\left(
        \frac{\beta_s m_0}{\beta_\tau}+a\xi
        \right)\right).
\label{hyper}
\end{equation}
Hence the maximum overlap is achievable for any temperature
below $(T_{m})_{\max}$.
The zero-temperature restoration (the MAP estimate)
is potentially as optimal as the finite temperature procedure
determined by Eq. (\ref{bound1}) in mean-field systems,
although it can only be achieved at the correct ratio $h/\beta_m$
given by Eq. (\ref{hyper}).

Figure \ref{fig_line} shows the line of optimal performance
for the parameters used in Fig. \ref{fig_IR1}.
The hyperbolas $T_m h$ = 0.9, 0.81 and 0.99 correspond to
the lines of operation Opt $h$, Opt*0.9 and Opt*1.1
of Fig. \ref{fig_IR1} respectively.
Where they intersect the line of optimal performance,
the overlap reaches a maximum.

In realistic image restoration, 
the precision in the estimation of hyperparameters 
is an important issue. 
One can see that the three lines of operation 
follow the general trend of the optimal curve. 
Hence they are much more error-tolerant than other curves, 
say, $h=1$.
Furthermore, if a line of operation 
intersects the line of optimal performance 
with a small angle between the tangents, 
then the overlap $M$ is very near to its optimum 
for a wide range of parameters along the line of operation, 
and the procedure has a high tolerance for parameter uncertainties. 
Among the three lines of operation in Fig. \ref{fig_line}, 
$T_m h$ = 0.99 has the highest tolerance. 
In fact, if one uses $T_m h$ = 1.0267 according to
Eq. (\ref{hyper}), then it has the widest range of tolerance
in the low temperature region. 

Figure \ref{fig_IR2} shows the effects of 
introducing the exchange term. 
It depicts $M$ as a function of the inverse exchange temperature
$\beta$ in Eq. (\ref{IRZ}) with the other parameters set
to the optimal values $T_s=T_m=0.9, a=h=1.0$ and with
$J_0=2.0$ in the unit $J=\tau=1$.
The axis $\beta =0$ corresponds to the optimum point $T_m=0.9$
of Fig. \ref{fig_IR1}.
The introduction of the exchange term is seen to
sharply improve the performance.
The maximum of $M$ is located at $\beta=2.0(=J_0)$ as required,
and $M$ stays close to the maximum value beyond $\beta=2.0$.
The combination of the ideas
of error-correcting code (the $\beta$-term)
and image restoration (the $\beta_m$ and $h$ terms) leads to a
remarkable improvement in the quality of restored image.

Two remarks are in order in relation to Fig. \ref{fig_IR2}.
First, the amount of information conveyed by the set $\{J_{ij}\}$
may seem exceedingly large compared to that by $\{\tau_i\}$
because the number of elements in the former set is
$N(N-1)/2$ while it is $N$ in the latter.
This fact may be mistaken as the reason of the improved result in
Fig. \ref{fig_IR2} for finite $\beta$.
However, since each $J_{ij}(\sim O(1/\sqrt{N}) )$ is
much smaller in magnitude than $\tau_i(\sim O(1))$, 
the contribution of each $J_{ij}$ is very small.
Such a situation is characteristic of the infinite-range model.
The equivalent situation in the finite-dimensional case
is that the number of exchange interactions is of the same
order as that of sites.
For example, there are $2N$ nearest neighbor interactions 
for $N$ sites on the square lattice.
Therefore, the increase in the amount of information
by the introduction of the set  $\{J_{ij}\}$ should be of
order unity, not infinitely large.
%
%
Secondly, as the exchange term is seen to increase the overlap very
sharply, even the information from a fraction of the exchange interactions
may be useful to improve the restoration result.
For example, one may choose a small fraction of pairs of sites
(either randomly or not) and use the corrupted, noisy version of these
exchange interactions to restore the image to obtain a better
result.
This method should be useful when the bandwidth (the amount of
information to be carried by the channel) is limited.
%

\subsection{Simulations}

It is difficult to investigate the 
more realistic case of two-dimensional images by
analytical methods.
We therefore have carried out Monte Carlo simulations
to confirm the qualitative pictures obtained by the
exact solution of the infinite-range model.
To generate the source image,
we have used the prior (\ref{ps}) with $T_s=2.15$
which is slightly lower than the critical point $2.269$
of the two-dimensional Ising model on the square lattice.
The error probability was set to $p_J=p_\tau=0.1$ for BSC,
corresponding to $\beta_J=\beta_\tau=1.0986$ by Eq. (\ref{betar}).
Averages over 5 samples (Fig. \ref{fig_IR3},
size $400\times 400$) or
10 samples (Fig. \ref{fig_IR4}, size $100\times 100$)
were taken at each data point.

Figure \ref{fig_IR3} shows the overlap $M$ as a function of
$T_m$ when $\beta=0$ and $h$ is chosen so that $h/\beta_m$ is
fixed to the optimum value $\beta_\tau /\beta_s$.
The overlap should have a maximum at $T_m=T_s=2.15$
in Fig. \ref{fig_IR3}
according to Eq. (\ref{bound1}) although it is not
very clearly seen due to statistical uncertainties.
It is at least true that $M$ does not change significantly
below $T_m=2.15$.
It is therefore unnecessary to lower the temperature than $T_s=2.15$
to obtain a better result.

The effects of exchange interactions have been taken into account
in Fig. \ref{fig_IR4} where $T_m$ and $h$ are fixed to the optimal
values 2.15($=T_s$) and 1.0986(=$\beta_\tau$) respectively.
The overlap is seen to increase quite significantly as a function
of $\beta$.
The axis $\beta=0$ corresponds to the optimal point of $T_m=2.15$
in Fig. \ref{fig_IR3}.
The overlap reaches its maximum at around $\beta=\beta_J=1.0986$
as it should and decreases slowly as $\beta$ is further increased.

Respectively, Figs. \ref{fig_IR3} and \ref{fig_IR4}
are qualitatively similar to Opt $h$ in Fig. \ref{fig_IR1},
and Fig. \ref{fig_IR2}, for the infinite-range model,
implying the usefulness of the infinite-range model as an
approximation of the two-dimensional problem.

Let us show an explicit example of the actual image restoration.
Figure \ref{fig_IR5} represents the situations of
Figs. \ref{fig_IR3} and \ref{fig_IR4} with the size $100\times 100$.
We have generated a pattern by the prior (\ref{ps})
with $T_s=2.15$ to obtain (a)
and have added noise with probability
$p_\tau=0.1$, resulting in (b) \cite{note}.
To obtain the restored images (c) and (d)
only the corrupted image (b) was used without extra
information on exchange interactions ($\beta =0$).
Restoration was tried at temperatures $T_m=0.5$ for (c)
resulting in $M=0.888$ and at $T_m=2.15$ for (d) with $M=0.892$.
It is clearly recognized that the optimal temperature $T_m=2.15$
(d) has a better restored image than (c).
The low-temperature process
(c) suppresses small structures
which were actually present in the original image.
The low temperature result is close to the MAP estimate
($T_m=0$) which would further suppress small structures.
It should be noticed that the difference in $M$ in these two restored
results (c) and (d) is very small (which is also seen in
Fig. \ref{fig_IR3}) but the intuitive impressions
on similarity to the original image (a) are rather different.
The reason is that the small structures do not contribute
significantly to the value of $M$ although such structures
have strong influence on intuitive impressions.
Therefore we should keep in mind that the overlap
$M$ alone does not
represent all aspects of the quality of restored images.

We next consider the effects of the additional information
of exchange interactions among nearest neighbors.
The same corrupted image (b)
has been used to obtain the restored image (e).
The parameters $T_m$, $h$ and $\beta$ were fixed
to the optimal values 2.15($=T_s$), 1.0986($=\beta_\tau$)
and 1.0986($=\beta_J$) respectively,
resulting in an overlap of $M=0.986$.
Fine structures are remarkably well restored in the result (e).
Thus the additional information of exchange interactions
is very effective to restore images faithfully.
%

\section{ERROR-CORRECTING CODES}

The infinite-range model has the same significance in error-correcting
codes as in the image restoration problem;
namely, an exactly solvable model which describes more realistic
situations at least qualitatively.
The difference is that we consider a
general value of $r$ in error-correcting codes, 
instead of only $r=2$ in the case of image restoration.
We therefore calculate the overlap
$M$ and related quantities explicitly assuming that
the set $\{i_1,\cdots,i_r\}$ in Eq. (\ref{Hamiltonian})
extends over all possible combinations of indices.

We consider the Gaussian channel, and the source and
model distributions are both assumed to be uniform,
$P_s=P_m=2^{-N}$, as is customary in the theory of error-correcting
codes \cite{Sourlas89,Rujan,Nishimori93,Sourlas94,Kabashima}.
>From Eqs. (\ref{M}) and (\ref{gauss1}), the overlap is given by
\begin{eqnarray}
  && M(\beta, h) \nonumber \\
  &=& 2^{-N} \sum_\xi 
    \int \prod dJ_{i_1 \cdots i_r}
    \left( \frac{N^{r-1}}{J^2 \pi r!}\right)^{1/2}
     \exp \left( -\frac{N^{r-1}}{J^2 r!}
      \sum_{i_1<\cdots <i_r} (J_{i_1 \cdots i_r}-\frac{j_0 r!}{N^{r-1}}
          \xi_{i_1}\cdots \xi_{i_r})^2
       \right) \nonumber \\
         &&\times 
     \int \prod d\tau_i \frac{1}{(\sqrt{2\pi}\tau )^N}
      \exp \left( -\frac{1}{2\tau^2}
       \sum_i (\tau_i -a \xi_i )^2 \right)
       \nonumber \\
   &&\times  \xi_i
   \, {\rm sgn}\left(
    \frac{\sum \sigma_i \exp \left(
         \beta \sum J_{i_1 \cdots i_r}\sigma_{i_1}\cdots \sigma_{i_r}
         +h \sum \tau_i \sigma_i \right)}
         {\sum \exp \left(
         \beta \sum J_{i_1 \cdots i_r}\sigma_{i_1}\cdots \sigma_{i_r}
         +h \sum \tau_i \sigma_i \right)}
         \right) .
         \label{ECC}
\end{eqnarray}
The normalizations of $J$ and $j_0$ are different from
Eq. (\ref{gauss1}) and follow the convention of the infinite-range
model of spin glasses so that the
limit $r\to\infty$ yields meaningful results \cite{Fischer}.

We may change the signs of integration variables in Eq. (\ref{ECC})
appropriately 
($J_{i_1 \cdots i_r}\to J_{i_1 \cdots i_r}\xi_{i_i}\cdots \xi_{i_r}$,
$\tau_i\to \tau_i\xi_i$, $\sigma_i\to \sigma_i\xi_i$)
which allows us to drop $\xi$'s from the integrand
(the ferromagnetic gauge).
Then the problem becomes the standard mean-field theory of
spin glasses with $r$-spin interactions under external random fields,
and we can apply the well-established replica method \cite{Fischer}.
%
%
Standard replica calculations
under the replica-symmetric (RS) ansatz
lead to the following set of
equations of state for the spin glass order parameter $q$,
ferromagnetic order parameter $m$ and the overlap $M$:
\begin{eqnarray}
  q&=&\int Dx \tanh^2 G  
  \label{eccq}\\
  m&=&\int Dx \tanh G 
  \label{eccm}\\
  M&=& \int Dx \, {\rm sgn}\, G
\end{eqnarray}
where $Dx$ is the Gaussian measure and
\begin{equation}
  G=\left(\frac{r\beta^2 J^2 q^{r-1}}{2}+\tau^2 h^2\right)^{1/2}x
    +\beta j_0 r\, m^{r-1} +a h .
\end{equation}
The corresponding free energy is
  \begin{equation}
    f_{\rm RS}=-T\log 2 -\frac{\beta J^2}{4}+\frac{\beta J^2}{4}(1-r)q^r
    +\frac{\beta J^2}{4}rq^{r-1}
    +j_0 (r-1) m^r -\int Dx \log \cosh G .
  \end{equation}

The present system with $j_0=0$ is known to have a spin glass phase
with a single-step replica-symmetry breaking (1RSB)
at a low temperature when $r\ge 3$ \cite{Gardner}.
This spin glass phase with 1RSB is replaced by a full-step 
replica-symmetry breaking at a still lower temperature.
It is therefore necessary to study replica-symmetry breaking
solutions following Refs. \cite{Gardner,Gross}.
The stability condition of the RS solution (the AT line) is
found to be
 \begin{equation}
   \frac{2T^2 q^{2-r}}{r(r-1)J^2} > \int Dx\, {\rm sech}^4 G.
   \label{AT}
 \end{equation}
The free energy with 1RSB ($h=0$ for simplicity) is
 \begin{eqnarray}
   f_{\rm 1RSB}&=&-T\log 2 -\frac{\beta J^2}{4}+
   \frac{\beta J^2}{4} x_0(1-r) q_0^r
    +\frac{\beta J^2}{4}(1-x_0)(1-r) q_1^r
    +\frac{\beta J^2}{4} rq_1^{r-1} 
    \nonumber\\
    &&   +j_0 (r-1) m^r
     -\frac{T}{x_0} \int Du \log \int Dv \cosh^{x_0} G_1
  \label{f-1RSB}
 \end{eqnarray}
where
 \begin{equation}
   G_1=u\sqrt{\frac{r}{2}\beta^2 J^2 q_0^{r-1}}
     + v \sqrt{\frac{r\beta^2 J^2}{2}}\sqrt{q_1^{r-1}-q_0^{r-1}}
     +\beta j_0 r m^{r-1} .
 \end{equation}
The self-consistent equations for the order parameters
are obtained by extremization of Eq. (\ref{f-1RSB})
 \begin{eqnarray}
   q_0&=& \int Du \left( \frac{
    \displaystyle{\int Dv \cosh^{x_0} G_1 \tanh G_1}}
    { \displaystyle{\int Dv \cosh^{x_0} G_1}} \right)^2
    \\
   q_1&=& \int Du \frac{\displaystyle{
    \int Dv \cosh^{x_0} G_1 \tanh^2 G_1}}
    {\displaystyle{\int Dv \cosh^{x_0} G_1}}
    \\
   m&=& \int Du  \frac{\displaystyle{
    \int Dv \cosh^{x_0} G_1 \tanh G_1}}
    {\displaystyle{\int Dv \cosh^{x_0} G_1}} .
  \end{eqnarray}
We do not write out the explicit form for the equation of $x_0$
because the formula is not very instructive.
The AT stability of this 1RSB solution is
  \begin{equation}
    \frac{2q_1^{2-r}T^2}{r(r-1) J^2} >
      \int Du \frac{\displaystyle{
    \int Dv \cosh^{x_0-4} G_1 }}
    {\displaystyle{\int Dv \cosh^{x_0} G_1}} .
  \end{equation}

The phase diagram in the case of
$r=3$ and $h=0$ is shown in Fig. \ref{fig_ecc1}.
Retrieval is not possible unless the ferromagnetic phase is
at least locally stable.
The hatched region satisfies this condition.
The ferromagnetic phase is stable in the replica symmetric ansatz 
for sufficiently strong bias $j_0$ and high temperature $T$. 
For $T/J$ above and below 0.651, 
it is respectively replaced by the paramagnetic and spin glass phases 
through first-order phase transitions when $j_0$ decreases. 
These three phases coexist at the triple point (TP).
The ferromagnetic phase remains metastable 
down to the spinodal line shown as a dotted curve. 
The replica symmetric solution of the ferromagnetic phase 
becomes unstable below the AT line (\ref{AT})
shown by the dash-dotted curve.

Also shown in Fig. \ref{fig_ecc1} are the spin glass phases, 
which exist at lower values of the bias $j_0$.
Spin glass with a single-step replica symmetry breaking 
is stable for $T/J$ between 0.651 and 0.240. 
At lower temperatures,
it is replaced by a full replica symmetry breaking spin glass phase. 

Investigation of the properties of the mixed phase M
(such as distinction between 1RSB and full RSB),
as well as the deatils of the spin glass phase,
are interesting future problems, which we do not pursue
here since they are not directly relevant to our problem
of error-correcting codes around the optimum temperature $T=J^2/2j_0$
shown dashed in Fig. \ref{fig_ecc1}.
We have not shown the structure of the phase diagram
at very low temperatures for this reason.

Figure \ref{fig_ecc2} shows the dependence
of the overlap $M$ on the decoding temperature $T=\beta^{-1}$
with $r=3$ and $j_0/J=0.77$.
The line $j_0/J=0.77$ lies slightly to the right of the triple
point TP in Fig. \ref{fig_ecc1}.
The maximum performance is achieved at $T=J^2/2j_0=0.649$.
This result is consistent with the argument in Sec. II C:
if we repeat the proof of Eq. (\ref{bound1}) with the external
field neglected and $P_s=P_m=2^{-N}$,
we obtain the inequality $M(\beta )\le M(2j_0/J^2)$.

The optimum condition $T=J^2/2j_0$ coincides with the Nishimori line
shown dashed in the phase diagram (Fig. \ref{fig_ecc1}).
This curve crosses the phase boundary and the spinodal line 
at the points where $j_0$ takes the smallest values 
in the ferromagnetic stable and metastable phases respectively 
(marked by black and white circles).
It can be shown that the spinodal line, the Nishimori line and the AT line 
are concurrent for any values of $r$. 
Hence the AT line terminates at the triple point. 
Since the Nishimori line lies in the replica symmetric phase, 
the replica symmetric argument would be sufficient
to clarify the behavior of the overlap around its maximum.
The lower temperature properties, including the possibility
of a reentrant spin glass phase,
may be affected by replica symmetry breaking.

Figure \ref{fig_ecc3} shows the dependence of $M$ 
on the random field strength $h$ at
the optimal temperature $T=0.649$ with
$r=3$, $j_0/J=0.77$ and $a=1$.
The axis $h=0$ corresponds to the conventional Sourlas code
without the field term,
which is the maximum point in Fig. \ref{fig_ecc2}.
It is observed that the overlap $M$ increases sharply
as the field is introduced, reaching the maximum at $h=1$,
in agreement with the theoretical prediction
$h_{\rm opt}=a/\tau^2$.
%

\section{DISCUSSIONS}

We have formulated the problems of image restoration and
error-correcting codes in a unified framework using
statistical mechanics.
We have derived an upper bound
on the overlap $M$ between the restored/decoded image/sequence
and the original image/sequence.
The maximum of $M$ is achieved when the restoring/decoding
temperature and field strength 
match the corresponding temperature and field strength 
characteristic of the source and channel properties.
This result comes as a natural generalization of the previously
known inequalities for image restoration \cite{Marroquin}
and error-correcting codes \cite{Nishimori93}.
The formulation and the proof of the inequality have a formal
similarity to the theory of spin glasses, in particular
the one using gauge symmetry \cite{Nishimori81}.
One should note however that we have not used gauge symmetry
in the present paper.
The variables $\{\xi_i\}$ in Eq. (\ref{expectation}),
playing a central role  in the spin glass theory
\cite{Nishimori81}, come
naturally in the present problem whereas they emerged as
a result of gauge transformation in the spin glass theory.

The infinite-range model has been solved exactly both in the
image restoration and error-correcting code situations.
The results made it possible to reveal the dependence of the
overlap $M$ on various parameters.
Simulations for image restoration
have confirmed that the results for
the infinite-range model remain qualitatively valid in
two dimensions.

For image restoration in mean-field systems, we have found 
a line of optimal performance along which the overlap $M$ 
takes the same maximum value. 
The line contains the point of optimal performance 
predicted by the inequality (\ref{bound1}), 
but extends also to the zero-temperature limit. 
This indicates that optimal (or quasi-optimal) performance 
is far more accessible than previously thought. 
It remains to study the extent to which the picture 
is applicable to finite-dimensional systems 
where the mean-field theory is only approximate.
In this respect, it is interesting to note that 
a ridge of nearly optimal overlap has already been observed 
in early literature, such as Fig. 2 of \cite{Marroquin} 
and Fig. 8 of \cite{Pryce}.
Naturally, one is led to expect that a narrow but extended region 
of optimal (or near-optimal) performance spans the parameter space.

By comparing the optimal line and the operation lines 
on which $h/\beta_m$ is kept constant, 
we have studied the tolerance towards uncertainties 
in parameter estimation.
Apparently the zero-temperature restoration (the MAP estimate) 
is most robust.
Furthermore, if the MAP estimate is approached by simulated annealing, 
it may be more effective to consider rescaling the field strength 
while lowering the temperature at the same time.

However, we have a few remarks of caution about the MAP estimate.
(a) The zero-temperature restoration is optimal 
only when the correct ratio $h/\beta_m$ is used, 
which can only be found self-consistently in realistic situations; 
if the incorrect ratio is used, the performance will be sub-optimal. 
(b) The existence of the line of optimal performance 
in finite-dimensional systems remains an open issue. 
Simulations in two dimensions seem to show that the MAP estimate
is sub-optimal, although most likely it is still nearly optimal.
On the other hand, the optimal point predicted by Eq. (\ref{bound1}) 
is guaranteed to be the best in general cases.
(c) The present result applies to the equilibrium state of the system, 
and the dynamics remains an open issue. 
It may happen that the approach to equilibrium at a low temperature 
is much slower, or is more prone to being trapped by local minima.

We have also considered the inclusion of exchange interactions 
as extra information in image restoration.
Explicit examples of images in two dimensions 
show that the fine structures are remarkably well restored. 
We remark that the exchange interactions have some similarities 
with ``line processes'', 
which has been proposed to improve the quality of images \cite{Geman}. 
If the line variables were quenched, 
they are equivalent to binary and multiple interactions 
among neighboring sites. 
However, a major difference is that the line variables are dynamical 
in the process of image restoration, 
whereas the exchange interactions considered here are quenched.

A comment is in order on the amount of information carried by the
channel of the infinite-range model.
The signal amplitude of the exchange term in Eq. (\ref{ECC})
is $j_0 r!/N^{r-1}$.
The channel noise causes fluctuations in the output with the
standard deviation $J(r!/N^{r-1})^{1/2}$, which is much
larger than the signal itself when $N\gg 1$.
This corresponds to an extremely low signal-to-noise ratio, 
yet the output still contains significant information
of the original message.
This demonstrates the power of the infinite-range decoding scheme 
in extremely noisy situations, 
although in practice such extremes do not occur frequently.

Finally, we mention briefly the idea of
selective freezing \cite{Wong96}.
The Ising spins keep moving under thermal agitation when
we employ the process of finite-temperature restoration.
Some spins have smaller thermal fluctuations than the others,
resulting in larger local magnetic moments.
It may thus be interesting to fix (freeze) those relatively
stable spins to $\pm 1$ according to the sign of
$\langle \sigma_i\rangle$
and repeat the finite-temperature decoding/restoration process
for the other less stable degrees of freedom.
We call this idea the selective freezing, which turns out to
enhance tolerance against uncertainties in parameter estimation.
The details will be presented in a forthcoming paper
\cite{Wong98}.
%

\acknowledgments
One of the authors (H. N.) thanks Department of Physics,
Hong Kong University of Science and Technology,
for hospitality.
He is also indebted to Prof. Kazuyuki Tanaka for a comprehensive
tutorial on the theory of image restoration
and to Prof. Shun-ichi Amari for useful comments.
We thank Dr. Domenico Carlucci for an observation leading to the
discovery  of the line of optimal performance,
Dr. David Saad for drawing our attention to Ref. \cite{Pryce},
and Prof. David Sherrington for discussions.
This work is partially supported by a grant from the Research Grant
Council of Hong Kong.
%

\appendix
\section{the inequality for general decoding and restoration}

To prove the inequality (\ref{bound1}),
we first note that the argument of
the summation in the definition (\ref{M}) 
is bounded by its absolute value:
\begin{eqnarray}
 &&M(\beta , h, P_m) \nonumber \\
 &=& \sum_\xi \prod \int dJ F_r(J) \prod \int d\tau F_1(\tau)
     \exp \left(\beta_J \sum J_{i_1 \cdots i_r}\xi_{i_1}\cdots \xi_{i_r} 
   +\beta_\tau \sum \tau_i \xi_i \right)
   P_s(\{\xi\} )  
        \xi_i \, {\rm sgn} \langle\sigma_i\rangle
        \nonumber \\
 &\le& \prod \int dJ F_r(J) \prod \int d\tau F_1(\tau
   \left| \sum_\xi \xi_i  \exp \left(
         \beta_J \sum J_{i_1 \cdots i_r}\xi_{i_1}\cdots \xi_{i_r}
         +\beta_\tau \sum \tau_i \xi_i \right)
          P_s(\{\xi\} ) \right| ,
    \label{Mintm}
\end{eqnarray}
where $|{\rm sgn}\langle \sigma_i \rangle |$ has been replaced with 1.
Using the identity $|x|=x\, {\rm sgn}x$, we get
\begin{eqnarray}
        && M(\beta , h, P_m)
        \nonumber \\
        &\le& \sum_\xi \prod \int dJ F_r(J) \prod \int d\tau F_1(\tau)
   \left( \sum_\xi \xi_i  \exp \left(
         \beta_J \sum J_{i_1 \cdots i_r}\xi_{i_1}\cdots \xi_{i_r}
         +\beta_\tau \sum \tau_i \xi_i \right)
          P_s(\{\xi\} ) \right)
          \nonumber \\
          && \times ~~~
    {\rm sgn}\left(
        \frac{\sum_\sigma \sigma_i  \exp \left(
         \beta_J \sum J_{i_1 \cdots i_r}\sigma_{i_1}\cdots \sigma_{i_r}
         +\beta_\tau \sum \tau_i \sigma_i \right)
          P_s(\{\sigma\}) }
         {\sum_\sigma \exp \left(
         \beta_J \sum J_{i_1 \cdots i_r}\sigma_{i_1}\cdots \sigma_{i_r}
         +\beta_\tau \sum \tau_i \sigma_i \right)
          P_s(\{\sigma\}) }\right).
\end{eqnarray}
Thus the right hand side can be interpreted
as the average of the product of $\xi_i$ and 
${\rm sgn}\langle \sigma_i\rangle$ at the optimal parameter values
$\beta =\beta_J$, $h=\beta_\tau$ and $P_m=P_s$,
yielding Eq. (\ref{bound1}).
%

\section{the inequality for mean-field image restoration}

To derive the inequality (\ref{bound2}), 
we start with the definition (\ref{M}). 
Substituting Eqs. (\ref{general}), (\ref{ps}) and (\ref{pm}) 
we obtain, for $i=1$ in the average,
\begin{eqnarray}
        M(h, \beta_m) = &&\frac{1}{Z(\beta_s)}
        \prod_i\int d\tau_i F_1(\tau_i)
        \sum_\xi \xi_1
     \exp \left(\frac{\beta_s}{z} 
        \sum_{\langle ij\rangle}\xi_i\xi_j
           +\beta_\tau \sum_i \tau_i \xi_i \right)
        \nonumber \\
      &&~~~\times 
    {\rm sgn} \left(
         \sum_\sigma \sigma_1
         \exp \left(
         \frac{\beta_m}{z}
        \sum_{\langle ij\rangle}\sigma_i\sigma_j
           +h \sum_i \tau_i \sigma_i \right)
        \right).
    \label{Mir}
\end{eqnarray}
In the exponential argument of the Boltzmann factor containing $\{\xi\}$, 
$\xi_1$ only appears in the expression
$z^{-1}\beta_s\sum_{\langle 1j\rangle}\xi_1\xi_j+\beta_\tau\tau_1\xi_1$. 
Hence if we multiply and divide this expression 
by the partition function of $\{\xi_i\}$ excluding site 1, 
we have
\begin{eqnarray}
        \sum_{\xi_i}\xi_1
     \exp \left(\frac{\beta_s}{z} 
        \sum_{\langle ij\rangle}\xi_i\xi_j
           +\beta_\tau \sum_i \tau_i \xi_i \right)=
        &&{\sum_{\xi_i}}^{\backslash 1}
     \exp \left(\frac{\beta_s}{z} 
        {\sum_{\langle ij\rangle}}^{\backslash 1}\xi_i\xi_j
           +\beta_\tau {\sum_i}^{\backslash 1}\tau_i \xi_i \right)
        \nonumber \\
        &&\left\langle \sum_{\eta_i}\eta_1
     \exp \left(\left(\frac{\beta_s}{z} 
        \sum_{\langle 1j\rangle}\eta_j
           +\beta_\tau \tau_1\right) \eta_1 \right)
        \right\rangle_{H(\beta_s,\beta_\tau)^{\backslash 1}},
\end{eqnarray}
where $\langle\cdots\rangle_{H(\beta_s,\beta_\tau)^{\backslash 1}}$ 
represents the thermal average taken over the Hamiltonian 
with inverse temperature $\beta_s$
and random field strength $\beta_\tau$, 
excluding site 1.
Similar arguments can be applied to the argument of the sign function 
in (\ref{Mir}), yielding
\begin{eqnarray}
        M(h, \beta_m)
        &&= \frac{1}{Z(\beta_s)}
        \prod_{i\ne 1}\int d\tau_i F_1(\tau_i)
        {\sum_{\xi_i}}^{\backslash 1}
     \exp \left(\frac{\beta_s}{z} 
        {\sum_{\langle ij\rangle}}^{\backslash 1}\xi_i\xi_j
           +\beta_\tau {\sum_i}^{\backslash 1}\tau_i \xi_i \right)
        \int d\tau_1 F_1(\tau_1)
        \nonumber \\
        && 2\left\langle \sinh \left(\frac{\beta_s}{z} 
        \sum_{\langle 1j\rangle}\eta_j
           +\beta_\tau \tau_1\right)
        \right\rangle_{H(\beta_s,\beta_\tau)^{\backslash 1}}
        {\rm sgn}\left\langle \sinh \left(\frac{\beta_m}{z} 
        \sum_{\langle 1j\rangle}\sigma_j
           +h \tau_1\right)
        \right\rangle_{H(\beta_m,h)^{\backslash 1}}.
\end{eqnarray}
For mean-field systems, 
$\sum_{\langle 1j\rangle}\eta_j$ and $\sum_{\langle 1j\rangle}\sigma_j$ 
are self-averaging quantities \cite{pb}, 
and the thermal average of the hyperbolic sine functions 
can be replaced by a single function of the thermal averaged argument. 
Thus $M(h,\beta_m)$ reduces to
\begin{eqnarray}
        M(h, \beta_m)
        &&= \frac{1}{Z(\beta_s)}
        \prod_{i\ne 1}\int d\tau_i F_1(\tau_i)
        {\sum_{\xi_i}}^{\backslash 1}
     \exp \left(\frac{\beta_s}{z} 
        {\sum_{\langle ij\rangle}}^{\backslash 1}\xi_i\xi_j
           +\beta_\tau {\sum_i}^{\backslash 1}\tau_i \xi_i \right)
        \int d\tau_1 F_1(\tau_1)
        \nonumber \\
        && 2 \sinh \left(\frac{\beta_s}{z} 
        \sum_{\langle 1j\rangle}
        \langle\eta_j
        \rangle_{H(\beta_s,\beta_\tau)^{\backslash 1}}
           +\beta_\tau \tau_1\right)
        {\rm sgn} \sinh \left(\frac{\beta_m}{z} 
        \sum_{\langle 1j\rangle}
        \langle\sigma_j
        \rangle_{H(\beta_m,h)^{\backslash 1}}
           +h \tau_1\right).
\label{M3}
\end{eqnarray}
For mean-field systems with large valency, 
the averaging over the neighbors of site 1 
reduces to the disordered average.
Consider $[\langle\eta_j\rangle^{\backslash 1}]$, which is
the thermal and disordered average of $\eta_j$
taken over the Hamiltonian $H(\beta_s,\beta_\tau)^{\backslash 1}$, 
\begin{eqnarray}
        [\langle\eta_j\rangle^{\backslash 1}]=
        &&\frac{1}{Z(\beta_s)^{\backslash 1}}
        \prod_{i\ne 1}\int d\tau_i F_1(\tau_i)
        {\sum_{\xi_i}}^{\backslash 1}
     \exp \left(\frac{\beta_s}{z} 
        {\sum_{\langle ij\rangle}}^{\backslash 1}\xi_i\xi_j
           +\beta_\tau {\sum_i}^{\backslash 1}\tau_i \xi_i \right)
        \nonumber \\
        &&\frac{
        {\sum_{\eta_i}}^{\backslash 1}\eta_j
     \exp \left(\frac{\beta_s}{z} 
        {\sum_{\langle ik\rangle}}^{\backslash 1}\eta_i\eta_k
           +\beta_\tau {\sum_i}^{\backslash 1}\tau_i \eta_i \right)}{
        {\sum_{\eta_i}}^{\backslash 1}
     \exp \left(\frac{\beta_s}{z} 
        {\sum_{\langle ik\rangle}}^{\backslash 1}\eta_i\eta_k
           +\beta_\tau {\sum_i}^{\backslash 1}\tau_i \eta_i \right)}.
\end{eqnarray}
After canceling terms in the denominator and numerator, we arrive at
\begin{equation}
        [\langle\eta_j\rangle^{\backslash 1}]=
        \frac{1}{Z(\beta_s)^{\backslash 1}}
        \prod_{i\ne 1}\int d\tau_i F_1(\tau_i)
        {\sum_{\eta_i}}^{\backslash 1}\eta_j
     \exp \left(\frac{\beta_s}{z} 
        {\sum_{\langle ik\rangle}}^{\backslash 1}\eta_i\eta_k
           +\beta_\tau {\sum_i}^{\backslash 1}\tau_i \eta_i \right),
\end{equation}
which reduces to $[\langle\xi_j\rangle^{\backslash 1}]=m_0$, 
namely the magnetization in the prior distribution.
Similarly, $[\langle\sigma_j\rangle^{\backslash 1}]=m$, 
which is the magnetization in the model distribution.

Substituting these results,
and using the normalization of the probability
${\sum_{\xi}}^{\backslash 1} P_s(\{\xi_i\}^{\backslash 1})
P(\{\tau_i\}^{\backslash 1}|\{\xi_i\}^{\backslash 1})$,
Eq. (\ref{M3}) reduces to
\begin{equation}
        M(h, \beta_m)
        = \frac{2 Z(\beta_s)^{\backslash 1}}{Z(\beta_s)}
        \int d\tau F_1(\tau)
        \sinh \left(\beta_s m_0 
           +\beta_\tau \tau\right)
        {\rm sgn} \left(\beta_m m
           +h \tau\right).
\end{equation}
The rest of the proof is similar to Appendix A.
Noting that the integrand of $\tau$ is bounded by its absolute value, 
we have
\begin{equation}
        M(h, \beta_m)
        \le \frac{2 Z(\beta_s)^{\backslash 1}}{Z(\beta_s)}
        \int d\tau F_1(\tau)
        \sinh \left(\beta_s m_0 
           +\beta_\tau \tau\right)
        {\rm sgn} \left(\beta_s m_0
           +\beta_\tau \tau\right).
\end{equation}
The right hand side is the value of $M(h,\beta_m)$ 
when Eq. (\ref{line}) is satisfied, since in this case,
${\rm sgn}(\beta_s m_0+\beta_\tau \tau)
={\rm sgn}(\beta_m m+h \tau)$.
%


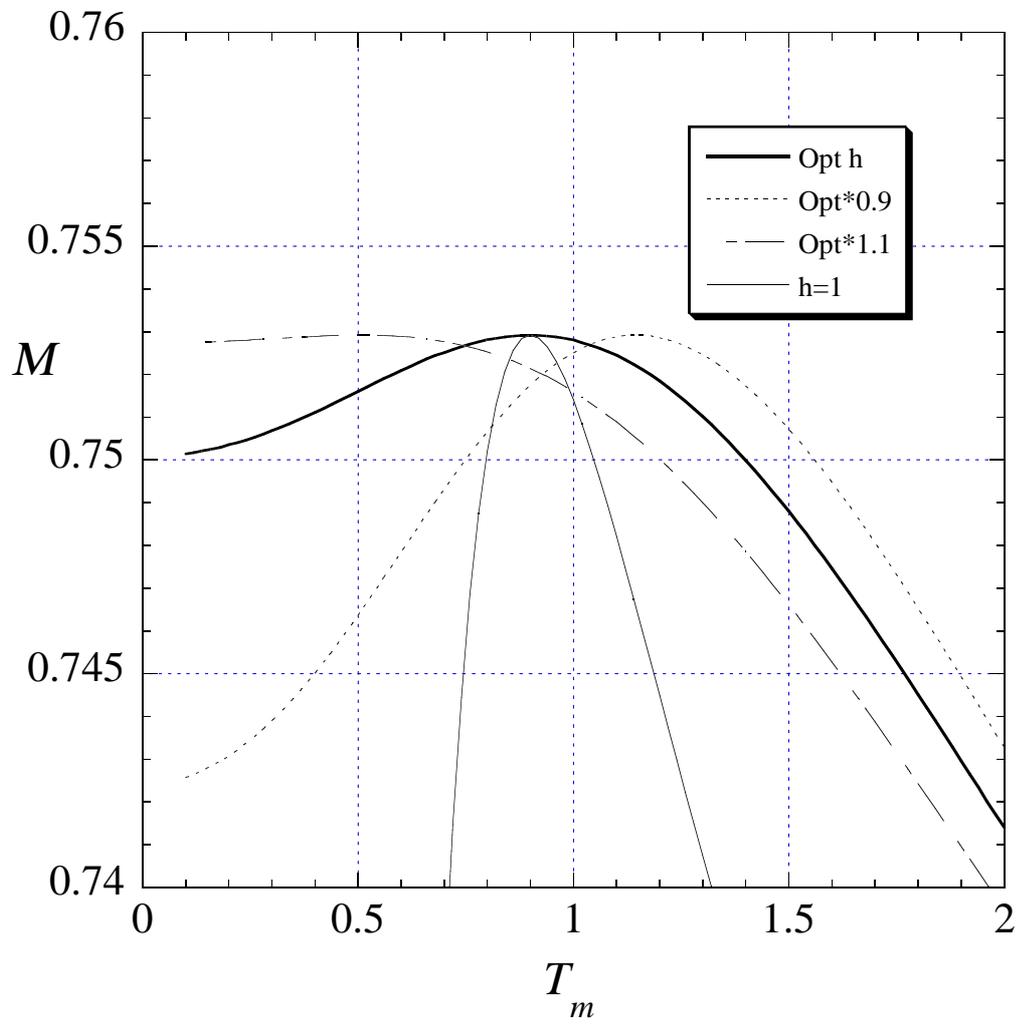
\begin{figure}
\caption{The overlap as a function of the restoration
temperature $T_m$ in the infinite-range model.
The random-field strength $h$ is chosen to be
$h=\beta_\tau\beta_m/\beta_s$ (Opt $h$),
$h=0.9\beta_\tau\beta_m/\beta_s$ (Opt*0.9),
$h=1.1\beta_\tau\beta_m/\beta_s$ (Opt*1.1),
or $h=1$.
}
\label{fig_IR1}
\end{figure}
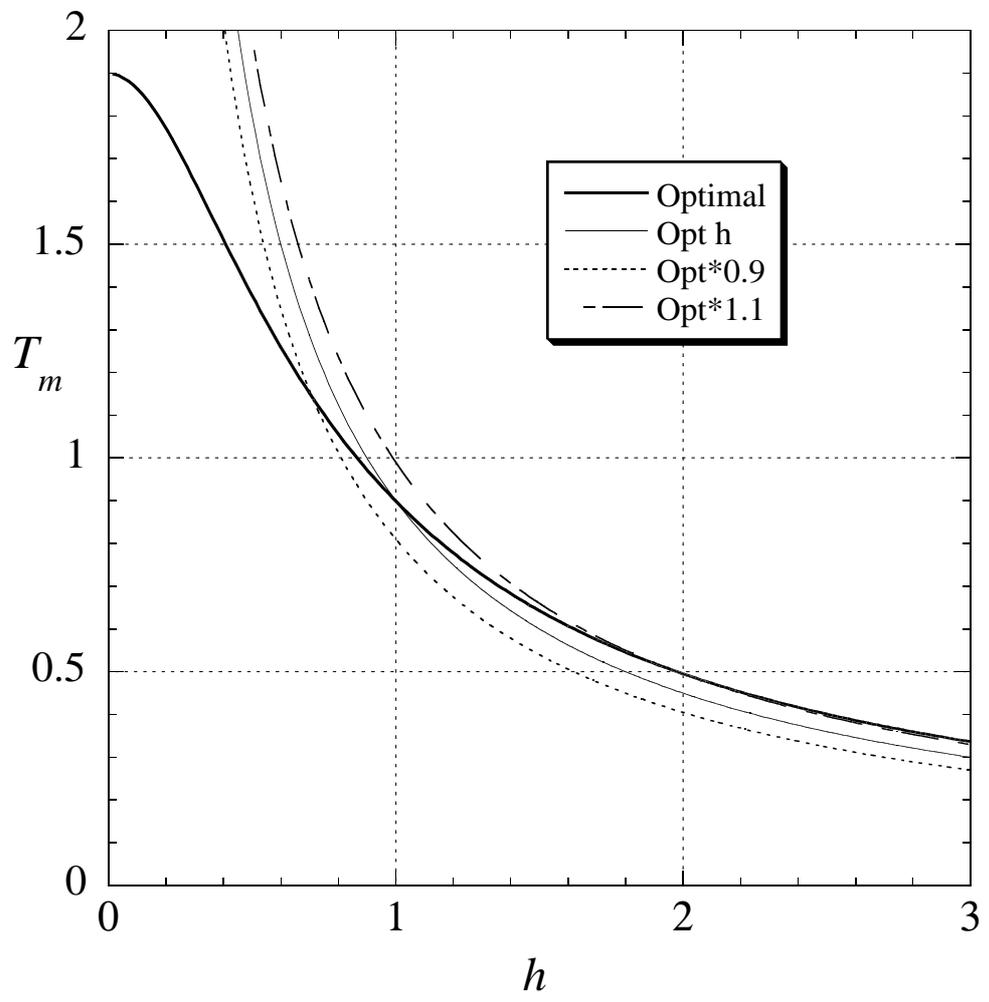
\begin{figure}
\caption{The line of optimal performance 
in the space of the random field strength $h$ 
and the restoration temperature $T_m$ 
in the infinite-range model 
for the parameters used in Fig. \ref{fig_IR1}. 
The three lines of operations 
(Opt $h$, Opt*0.9 and Opt*1.1) 
are shown for comparison.
}
\label{fig_line}
\end{figure}
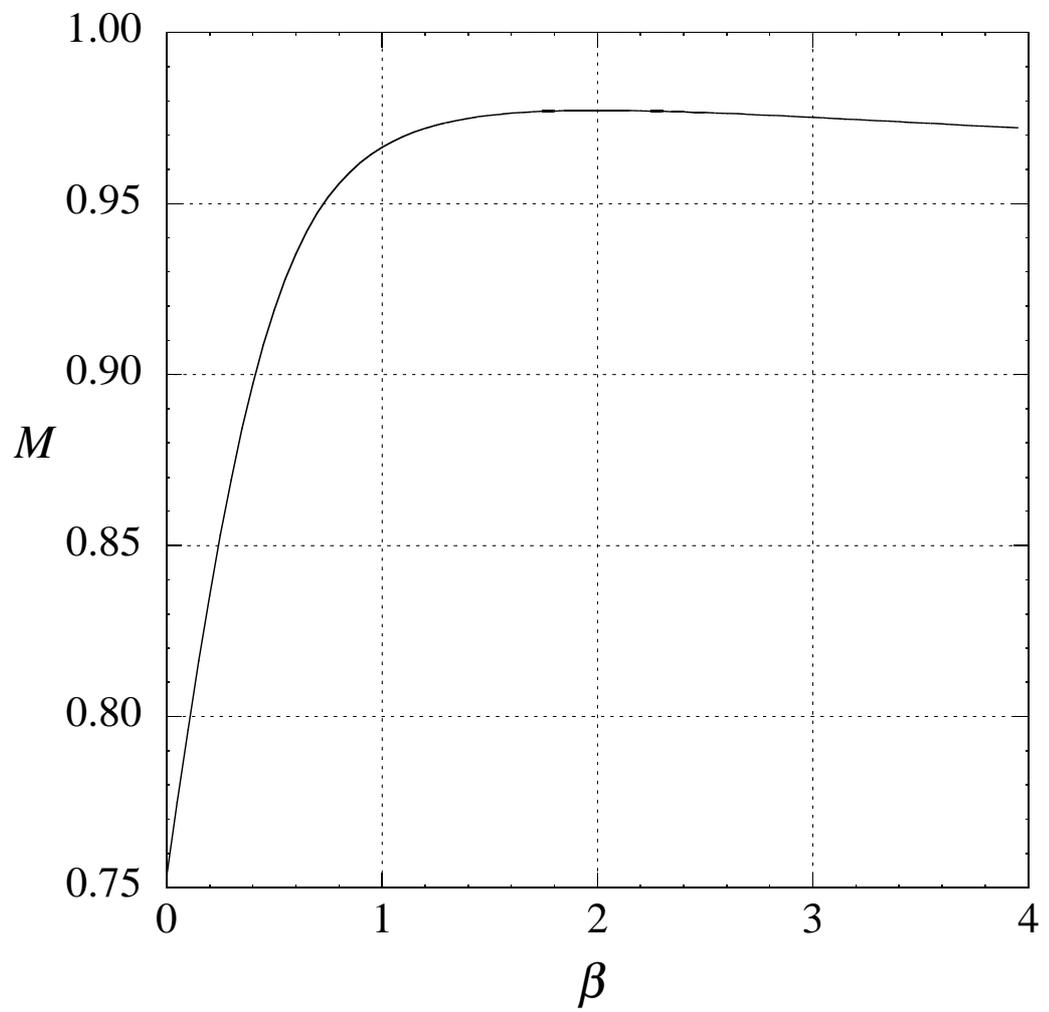
\begin{figure}
\caption{The overlap as a function of the parameter $\beta$
in the infinite-range model.
The center of the channel output distribution is $J_0=2.0$ and
the other restoration parameters are chosen to be the optimal
values $T_m=T_s(=0.9)$ and $h=a(=1)$.
%
}
\label{fig_IR2}
\end{figure}
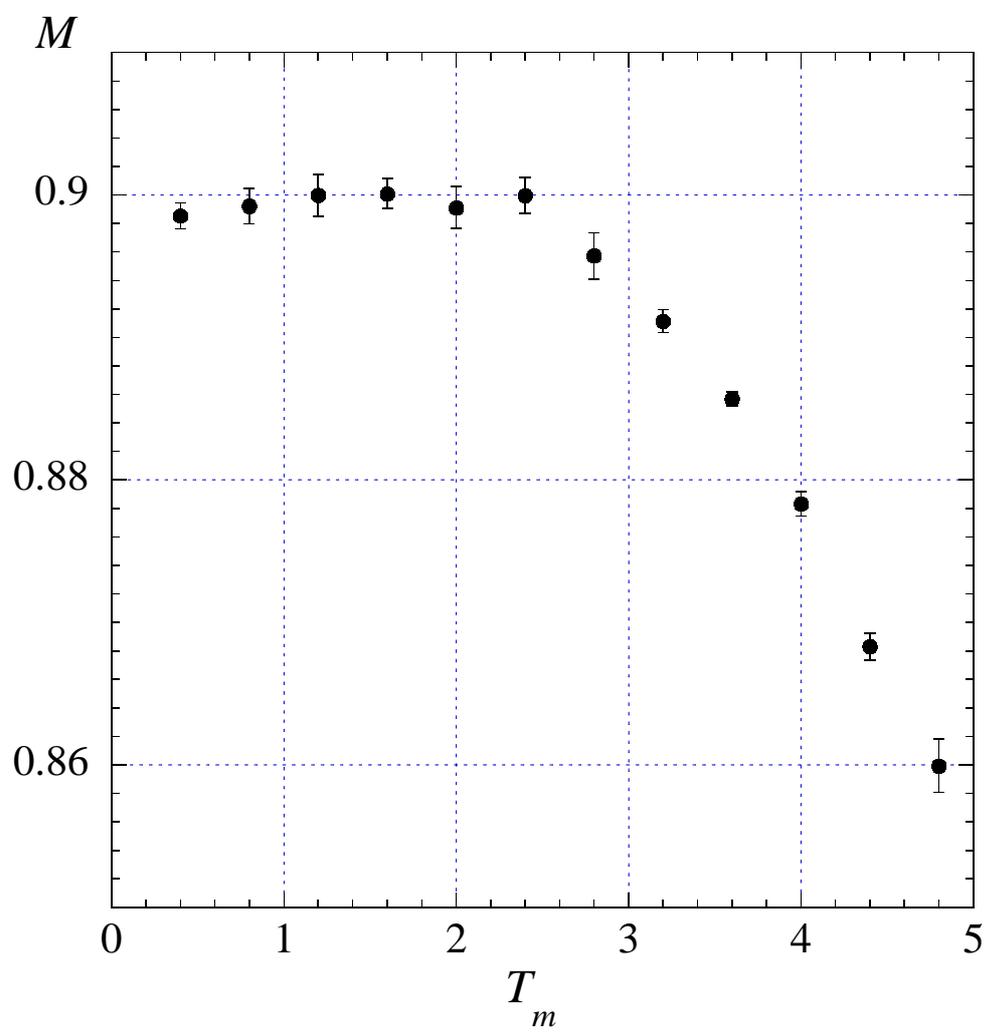
\begin{figure}
\caption{The overlap as a function of the restoration temperature
on the square lattice obtained by simulations.
The source temperature is $T_s=2.15$, the error
rate $p_\tau=0.1$, and
no exchange interactions ($\beta=0$).
}
\label{fig_IR3}
\end{figure}
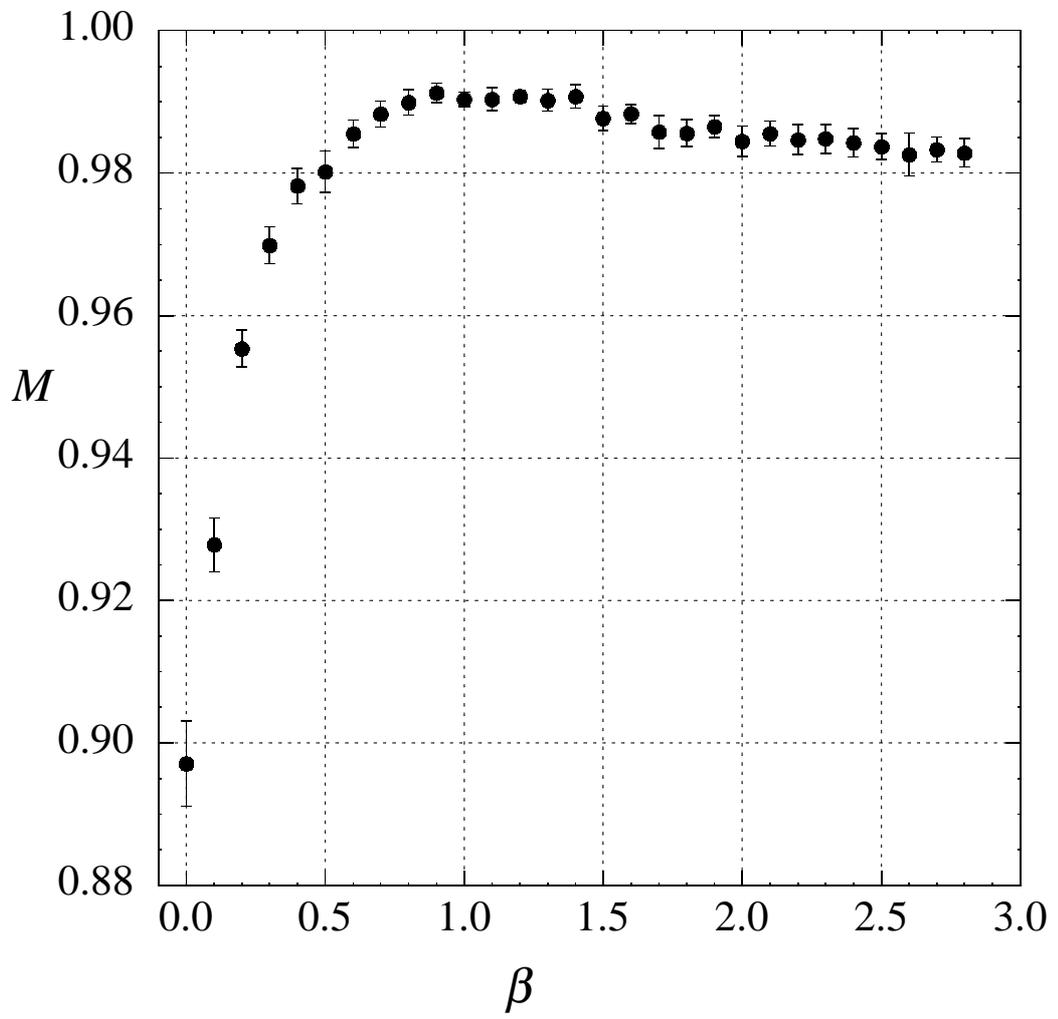
\begin{figure}
\caption{The overlap as a function of $\beta$
for the square lattice when $T_m=T_s=2.15$ and
$h=\beta_\tau=1.0986$.
%
%
}
\label{fig_IR4}
\end{figure}
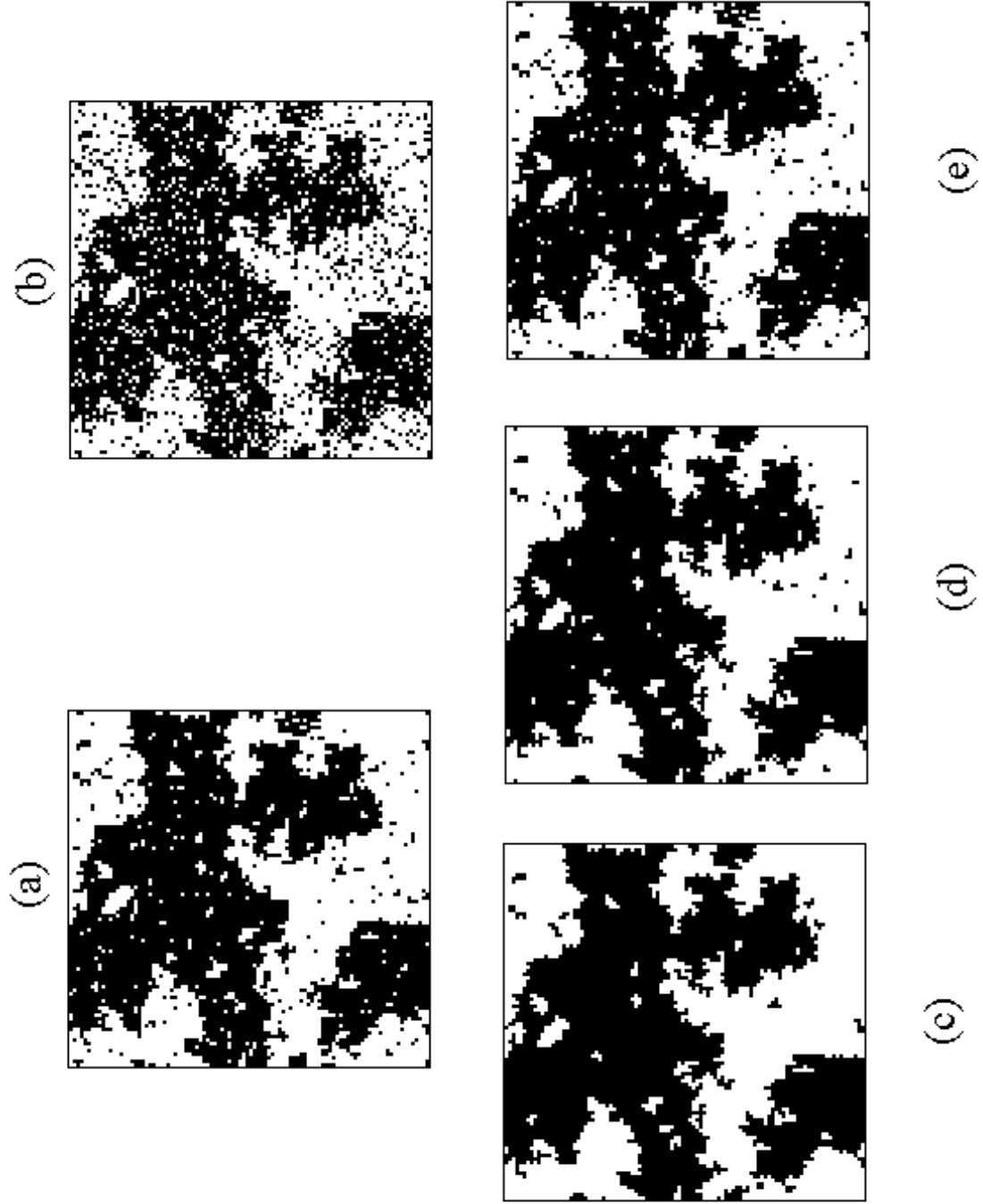
\begin{figure}
\caption{
Examples of image restoration.
The original image is (a), and the image corrupted
by the noise ($p_\tau=0.1$) is (b).
The restored images are (c) ($T_m=0.5$), (d) ($T_m=2.15$),
and (e) (with the exchange term).
}
\label{fig_IR5}
\end{figure}
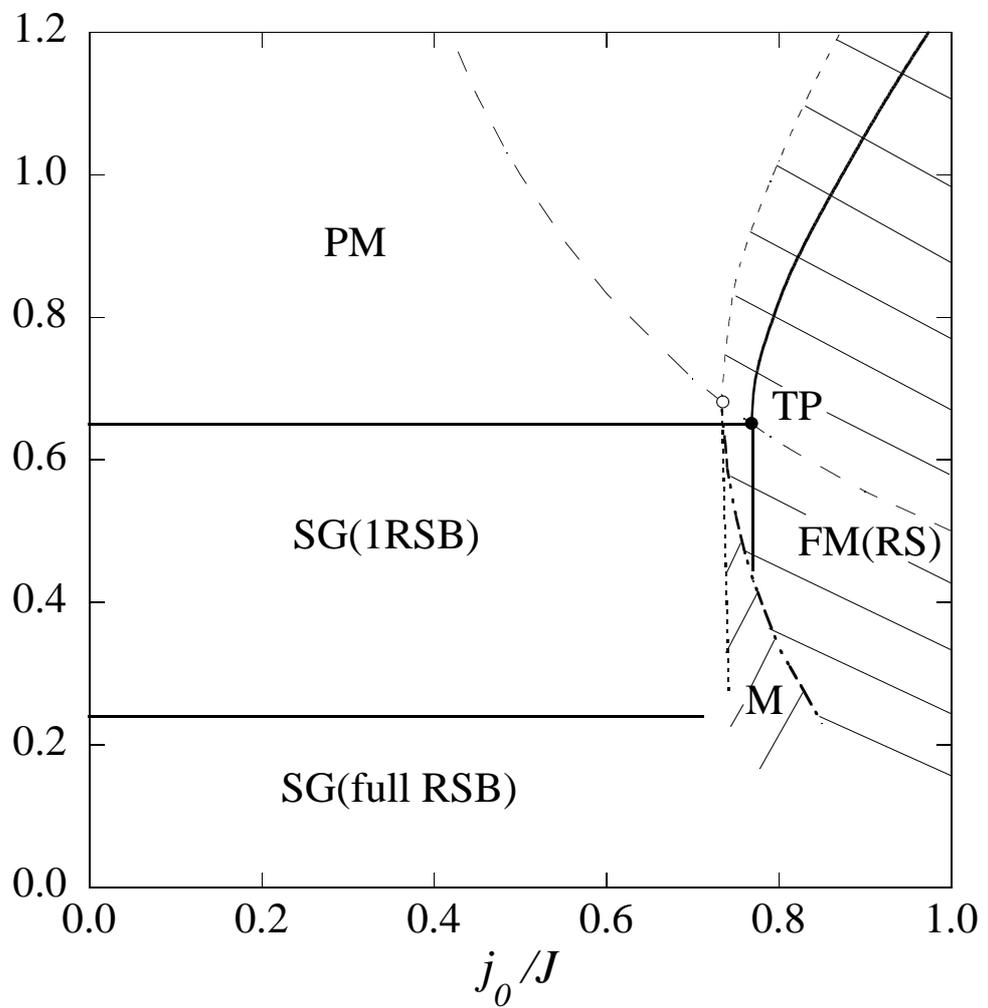
\begin{figure}
\caption{The phase diagram of the $r=3$ system.
Message retrieval is possible in the stable and metastable
ferromagnetic phases shown hatched.
}
\label{fig_ecc1}
\end{figure}
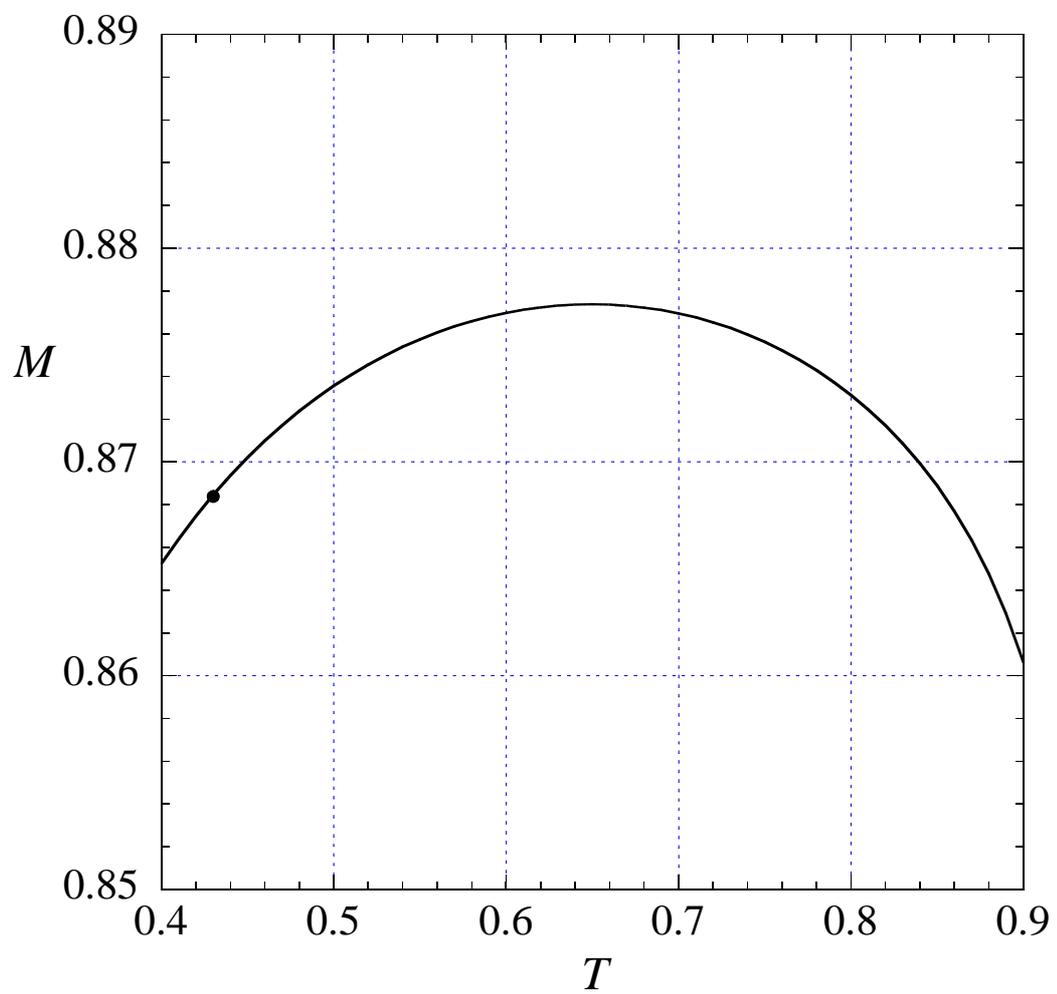
\begin{figure}
\caption{The overlap $M$ as a function of the decoding temperature
in the Sourlas code.
 Three-body interactions $r=3$ are considered and
the center of the channel output distribution is $j_0/J=0.77$.
The field term is $h=0$.
The replica-symmetric solution shown here is unstable below
the AT line at $T=0.43$ (shown dotted) although we do not
expect a significant deviation in the temperature
range $T\ge 0.40$.
%
}
\label{fig_ecc2}
\end{figure}
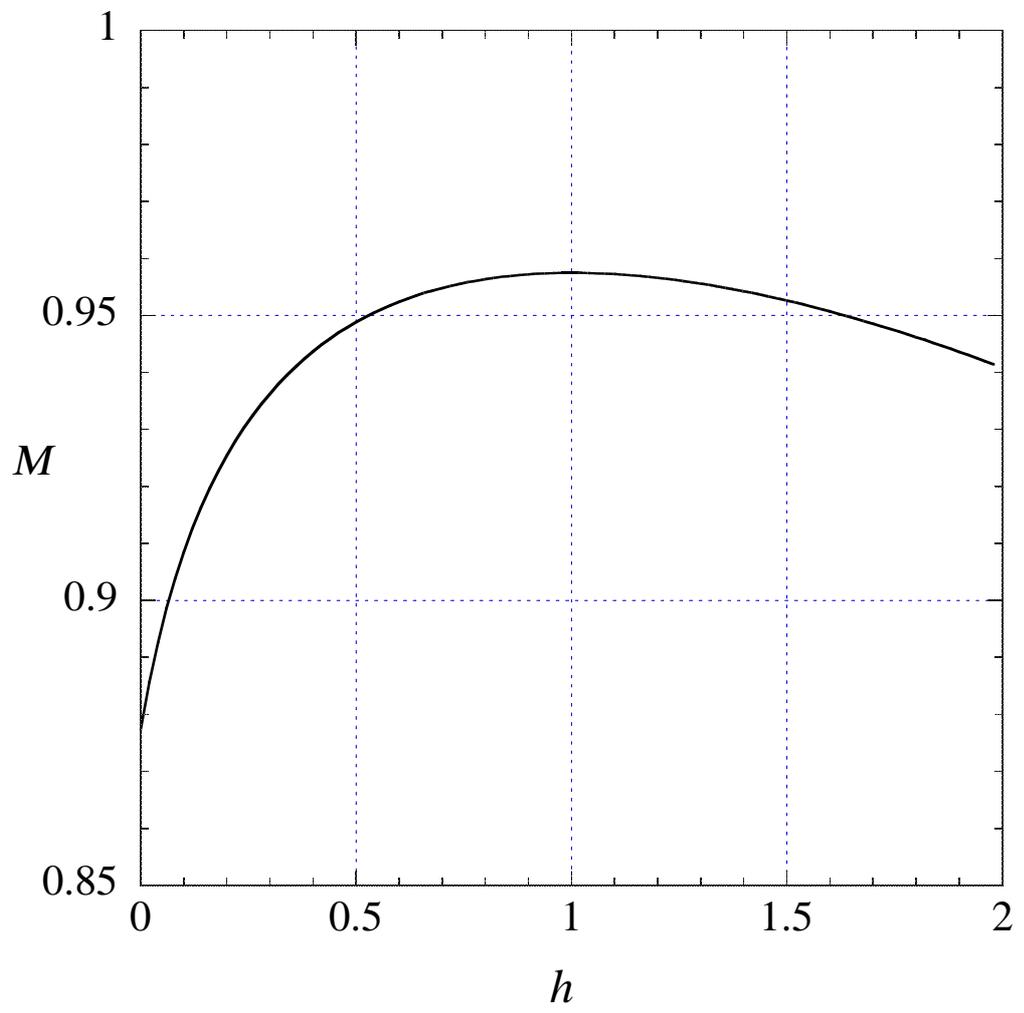
\begin{figure}
\caption{The overlap as a function of the field strength $h$.
The conventional statistical-mechanical formulation of
error-correcting codes (Sourlas code) corresponds
to the axis $h=0$.
The parameters are $r=3, j_0/J=0.77, a=1$ and $T=0.649$.
%
}
\label{fig_ecc3}
\end{figure}

\begin{thebibliography}{99}
\bibitem{note}
Rigorously speaking, the state of Fig. \ref{fig_IR5} (a)
is a metastable state, not in full thermal equilibrium in
which most of the pixels are either black or white
when $T_s$ is below the critical point.
%
We nevertheless show the example (a) in Fig. \ref{fig_IR5}
because it represents true equilibrium situations
rather faithfully (if we look only at the
mostly white domain, for example) and simultaneously
it is easier to intuitively recognize the differences among
(a)-(e) than by using mostly black or white images.
%
The simulations of Figs. \ref{fig_IR3} and \ref{fig_IR4} were
carried out in true equilibrium.
\end{thebibliography}
\end{document}